\documentclass[a4paper, amsfonts, amssymb, amsmath, reprint, showkeys, nofootinbib]{revtex4-1}
\usepackage[english]{babel}
\usepackage{amsmath,amssymb,amsthm}
\usepackage{mhchem}
\usepackage[utf8]{inputenc}
\usepackage{xcolor}
\usepackage{amsthm}
\usepackage{mathtools}
\usepackage{palatino}
\usepackage{physics}
\usepackage{xcolor}
\usepackage{libertine} 
\usepackage{changepage}
\usepackage{graphicx}
\usepackage[left=20mm,right=20mm,top=30mm,columnsep=15pt]{geometry} 
\usepackage{adjustbox}
\usepackage{placeins}
\usepackage[T1]{fontenc}
\usepackage{lipsum}
\usepackage{csquotes}
\usepackage{float}
\usepackage{textcomp}
\usepackage[pdftex, pdftitle={Article}, pdfauthor={Author}]{hyperref} 
\usepackage{setspace}
\newcommand{\nocontentsline}[3]{}
\let\origcontentsline\addcontentsline
\newcommand\stoptoc{\let\addcontentsline\nocontentsline}
\newcommand\resumetoc{\let\addcontentsline\origcontentsline}

\newpage

\begin{document}
\title{End-to-end physics-based modeling of laser-activated color centers in silicon}

\author{Qiushi Gu$^{1,*}$, Valeria Saggio$^{1}$, Camille Papon$^{1}$, Alessandro Buzzi$^{1}$, Ian Christen$^{1}$, Christopher Panuski$^{1}$, Carlos Errando-Herranz$^{1,2,3}$, and Dirk Englund$^{1,}$}
\email{qiushigu@mit.edu, englund@mit.edu}

\address{\vspace{0.2cm} $^1$Massachusetts Institute of Technology, Cambridge, MA, USA\\
$^2$QuTech and Kavli Institute of Nanoscience, Delft University of Technology, Delft, Netherlands\\
$^3$Department of Quantum and Computer Engineering, Delft University of Technology, Delft, Netherlands} 

\begin{abstract}
Color centers are among the most promising candidates for quantum information processing. Central requirements for their practical applications include controlled and efficient local activation in nanophotonic devices and identical spectral features. However, producing color centers in a controlled and reliable way is inherently challenging due to the lack of comprehensive theoretical insights into their formation and the difficulty of streamlining the generation process for rapid in-situ optimization. We address these challenges by developing an end-to-end first-principles model that captures the underlying formation process of color centers. Emitters are activated through laser annealing, which allows for in-situ creation and the possibility of model-based control. 
Notably, our model enables the estimation of the emitters' inhomogeneous broadening down to $\sim~16~\mathrm{GHz}$ in bare silicon, which translates into the creation of emitters with highly similar spectral properties. Finally, we address the challenge of in-situ deterministic activation of color centers in nanophotonic devices by going beyond bare silicon and demonstrating successful laser writing in photonic crystal optical cavities. These results lay the foundation for deterministic and large-scale integration of color centers within quantum photonic platforms.
\end{abstract}

\maketitle

\stoptoc

\section{Introduction}
Among the leading platforms for the development of quantum networks and distributed quantum information processing, color centers, particularly in silicon, emerge as prominent candidates~\cite{afzal2024distributed, simmons2024scalable}. A key advantage of these systems is their coherent spin-photon interfaces, which enable local manipulation of quantum information in their spin states and its transfer via photonic qubits~\cite{ruf_quantum_2021}. Additionally, the use of silicon as host material for these emitters offers a unique benefit in that it allows emission in the telecommunication wavelength band~\cite{udvarhelyi_identification_2021,redjem_single_2020,higginbottom_optical_2022} --- which is essential for long-distance quantum information transfer over optical fibers --- and leverages the maturity of the silicon industry to streamline the nanofabrication process and ensure compatibility with the state-of-the-art microelectronic and photonic technology~\cite{saggio2024cavity,komza2024indistinguishable,kim2024bright,prabhu2023individually,redjem2023all,lefaucher2023cavity}. \\

However, central challenges in this field include i) the local activation of color centers within their host material in a controlled and on-demand fashion, to enable rapid in-situ creation, ii) the extension of this method to activation in nanophotonic devices, to facilitate large-scale integration, and iii) the generation of emitters with spectral features closely aligned to one another, to enable high-quality multi-photon interference. While localized creation of emitters in implanted silicon-on-insulator (SOI) has been reported~\cite{jhuria2024programmable,andrini2024activation,quard2024femtosecond,hollenbach_wafer-scale_2022}, reliable generation is inherently difficult to predict and control. This is partly due to the limited theoretical understanding of the physics governing their formation and partly due to challenges in streamlining the process for rapid optimization of experimental generation parameters in situ. 

We address this challenge by developing an end-to-end, first-principles model (i.e. a digital twin) that captures the underlying creation mechanism of laser-annealed color centers. We demonstrate our method in silicon as a test case, noting that the approach also applies to color centers in other host materials such as diamond or silicon carbide. 
We benchmark our digital twin against experimental laser-annealing results, which allows us to gain novel physical insights into the properties of the generated color centers (G-centers in silicon in our case). For example, we obtain an estimate of their number, information about their excited-state lifetime, and activation and dissociation energies. Remarkably, our model also enables a more precise attribution of broadening mechanisms, leading to the extraction of a very narrow inhomogeneous broadening of our generated emitters with values as low as $\sim~88~\mathrm{pm}$ ($16~\mathrm{GHz}$) in bare silicon --- up to at least two orders of magnitude lower than previously reported for annealed emitters in bulk silicon~\cite{jhuria2024programmable, quard2024femtosecond}. This translates into color centers with closely matching spectral properties, which is essential for quantum information processing~\cite{waldermann2007creating}.

\begin{figure*}[t]
  {\centering
\includegraphics[width=\linewidth]{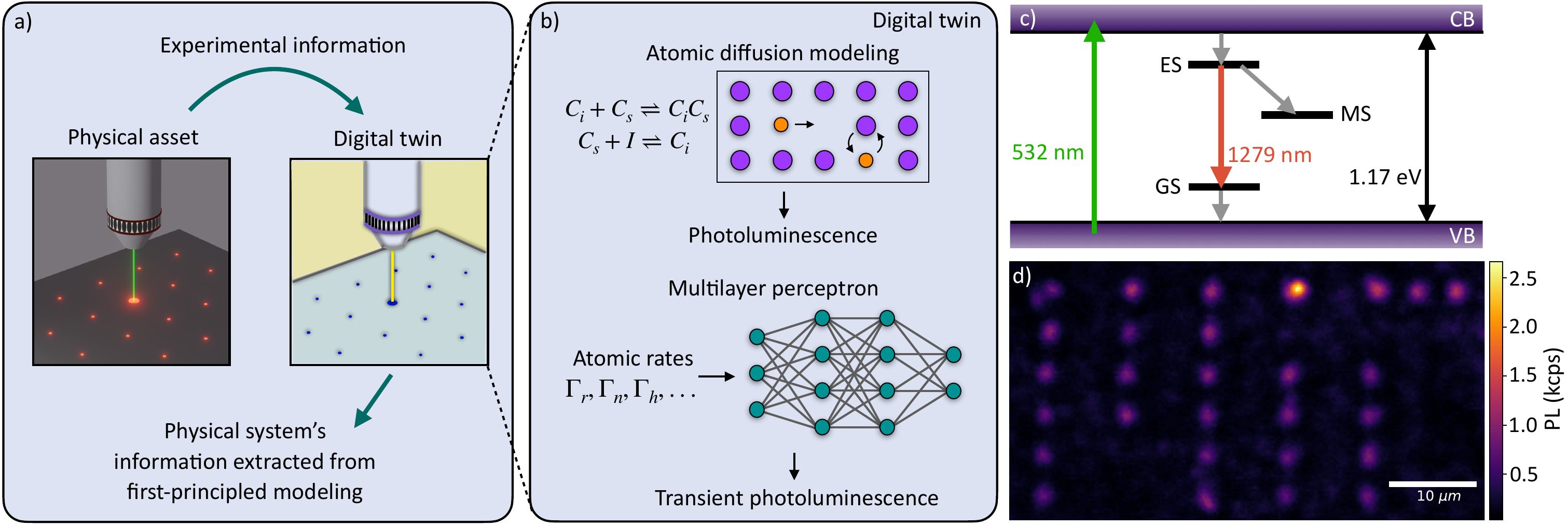}}
\caption{\textbf{Experimental annealing technique and theoretical modeling.} a) Schematic of our laser annealing method, where a green laser beam is focused through an objective onto pre-determined locations on the sample surface, along with its digital twin. The experimental information deriving from our physical asset is fed into our digital twin, which returns physical insights into the activated emitters (G-centers). b) Digital twin of the laser annealing process, incorporating atomic diffusion of carbon in silicon and a multilayer perceptron (MLP) that processes diffusion-related rates. The digital twin directly predicts the photoluminescence (PL) and transient PL. c) Energy levels of a G-center featuring a radiative transition between two singlet states GS and ES with emission in the telecom O-band ($\sim 1279~\mathrm{nm}$) and a non-radiative transition involving an additional triplet metastable state (MS). The electrons are promoted from the valence band (VB) to the conduction band (CB) upon excitation with laser light at $532~\mathrm{nm}$.
d) Laser-annealed MIT logo, where each spot corresponds to the PL generated by emitter activation.}
\label{fig1}
\end{figure*}

Lastly, we extend our laser annealing technique and its modeling to direct laser writing of color centers into photonic devices such as inverse-designed 2D optical cavities~\cite{panuski2022full,saggio2024cavity}, unlocking the potential for scalable integration. 

\section{Results}
Emitters are activated in our sample at predefined spots with the use of a continuous-wave (CW) visible ($532~\mathrm{nm}$) laser beam --- the annealing beam --- focused on the sample using an objective with a numerical aperture (NA) of $0.65$. An illustration of this laser-annealing technique is shown in the left inset of Fig.~\ref{fig1}a. 
The right inset of Fig.~\ref{fig1}a shows its digital twin, which uses experimental information derived from our physical asset to output novel insights about it. The structure of our digital twin is illustrated in Fig.~\ref{fig1}b. It is composed of two parts, one accounting for the atomic diffusion processes of the carbon atoms in silicon (\ce{C_i} and \ce{C_s} representing interstitial and substitutional atoms, respectively), and one using a Lindbladian master equation approximated by a multilayer perceptron (MLP) where the atomic rates of the system are used as input (see SI Sec.~\ref{digitaltwinSI} for details).

The created emitters are identified as G-centers~\cite{redjem_single_2020}, composed of two substitutional carbon atoms and a silicon interstitial. Their properties are characterized with a separate CW laser beam at $532~\mathrm{nm}$ --- the excitation beam --- 
which excites our emitters as shown in Fig.~\ref{fig1}c. The G-center features two singlet states --- ground (GS) and excited (ES) --- and an expected spin triplet metastable state (MS), observed in ensembles only so far~\cite{udvarhelyi2021identification,lee1982optical}. Upon $532~\mathrm{nm}$ excitation, electrons transition from the valence band (VB) to the conduction band (CB) before decaying to the GS, emitting photons at $1279~\mathrm{nm}$ (or $970~\mathrm{meV}$), in the telecom O-band. The emission at $1279~\mathrm{nm}$ is referred to as the zero-phonon line (ZPL). The annealing and excitation beams are directed onto the sample (kept at gryogenic temperatures) via scanning mirrors for precise positioning (details on the experimental apparatus in SI, Sec.~\ref{sec:setup}).
The digital twin modeling provides a way to directly compare the generated photoluminescence (PL) and transient PL with experimental results, without judicious yet subjective selection of experimental features to compare with phenomenological models. This comparison also enables the extraction of system parameters that govern emitter activation. An example of emitter formation via laser annealing on bare silicon is shown as a color map in Fig.~\ref{fig1}d, where the MIT logo has been reproduced by varying the annealing beam position while keeping its power and illumination duration constant. The visible spots correspond to the PL generated by the activated emitters.

\begin{figure*}
  {\centering
\includegraphics[width=\linewidth]{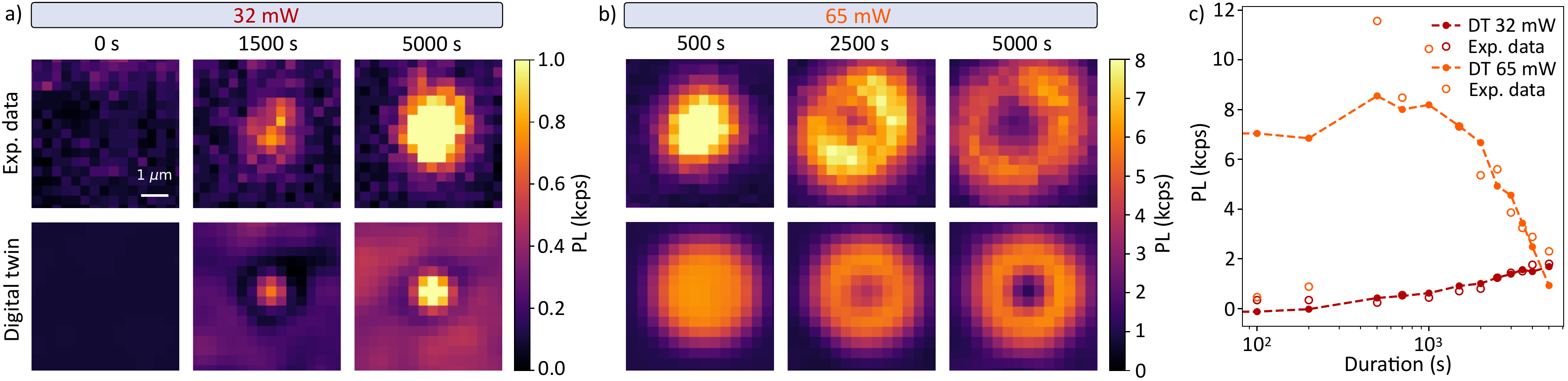}}
\caption{\textbf{Comparison between experimental and digital twin-modeled PL.} a) PL of laser annealed spots obtained from experiments (top row) and digital twin modeling (bottom row) at an annealing power of $32~\mathrm{mW}$ for different annealing durations ($0~\mathrm{s}$, $1500~\mathrm{s}$ and $5000~\mathrm{s}$). Emitters start to form after $1500~\mathrm{s}$. b) Experimental results (top row) and digital twin-predicted outcomes (bottom row) at $65~\mathrm{mW}$ for different annealing durations ($500~\mathrm{s}$, $2500~\mathrm{s}$ and $5000~\mathrm{s}$). In this case, emitters are erased with longer exposure times. The deactivation becomes visible around $2500~\mathrm{s}$. c) Experimental and digital twin-predicted PL counts at the central pixel of the annealing beam versus annealing duration at low ($32~\mathrm{mW}$) and high ($65~\mathrm{mW}$) annealing powers. DT in the legend denotes the digital twin data.}
\label{fig2}
\end{figure*}

The generation mechanism of the emitters is likely the thermal activation of an atomic diffusion process~\cite{davies1989annealing}. The elevated annealing-induced temperature promotes carbon diffusion via the Watkins' replacement reaction, leading to more frequent encounters of two carbon atoms and the formation of G-centers. At higher temperatures, G-center dissociation dominates and results in erasure of the formed emitters. We model this process with a digital twin based on atomic diffusion (see SI, Sec.~\ref{digitaltwinSI1}), and systematically study the effect of different annealing powers and durations on the same location. At $32~\mathrm{mW}$, the PL brightness (or the number of emitters) increases with annealing duration on a timescale of $\sim~1000~\mathrm{s}$, as shown in Fig.~\ref{fig2}a top row. This is consistent with the formation energy of G-centers, which is determined to be $E_a^f=(0.83\pm0.01)~\mathrm{eV}$ from our digital twin. When increasing the power to $65~\mathrm{mW}$, emitters first start to form and then get erased with further exposure, as visible in Fig.~\ref{fig2}b top row. This is consistent with the dissociation of G-centers at high temperatures, observed also in earlier experiments where rapid thermal annealing at $1000~\mathrm{^\circ C}$ results in the reduction of brightness of ensembles of G-centers and appearance of single G-centers~\cite{zhiyenbayev2023scalable, redjem2023all}, or in deactivation of G-centers~\cite{prabhu2023individually}. From our digital twin, we extract a dissociation energy of $E_a^b=(1.962\pm0.01)~\mathrm{eV}$ (see SI Sec.~\ref{digitaltwinSI1} for more details). Erased emitters cannot be restored with lower-power annealing. 
We extract the activation energies by comparing the experimental PL maps shown in Figs.~\ref{fig2}a and b top rows with the digital twin-generated PL maps, reported in Figs.~\ref{fig2}a and b bottom rows. Fig.~\ref{fig2}c displays the experimental PL values at the central pixel of the annealing beam along with the digital twin-predicted PL for both powers, showing a monotonic PL increase at lower powers and a drop after $\sim~1000~\mathrm{s}$ at higher powers.

\begin{figure*}
  {\centering
  \includegraphics[width=\linewidth]{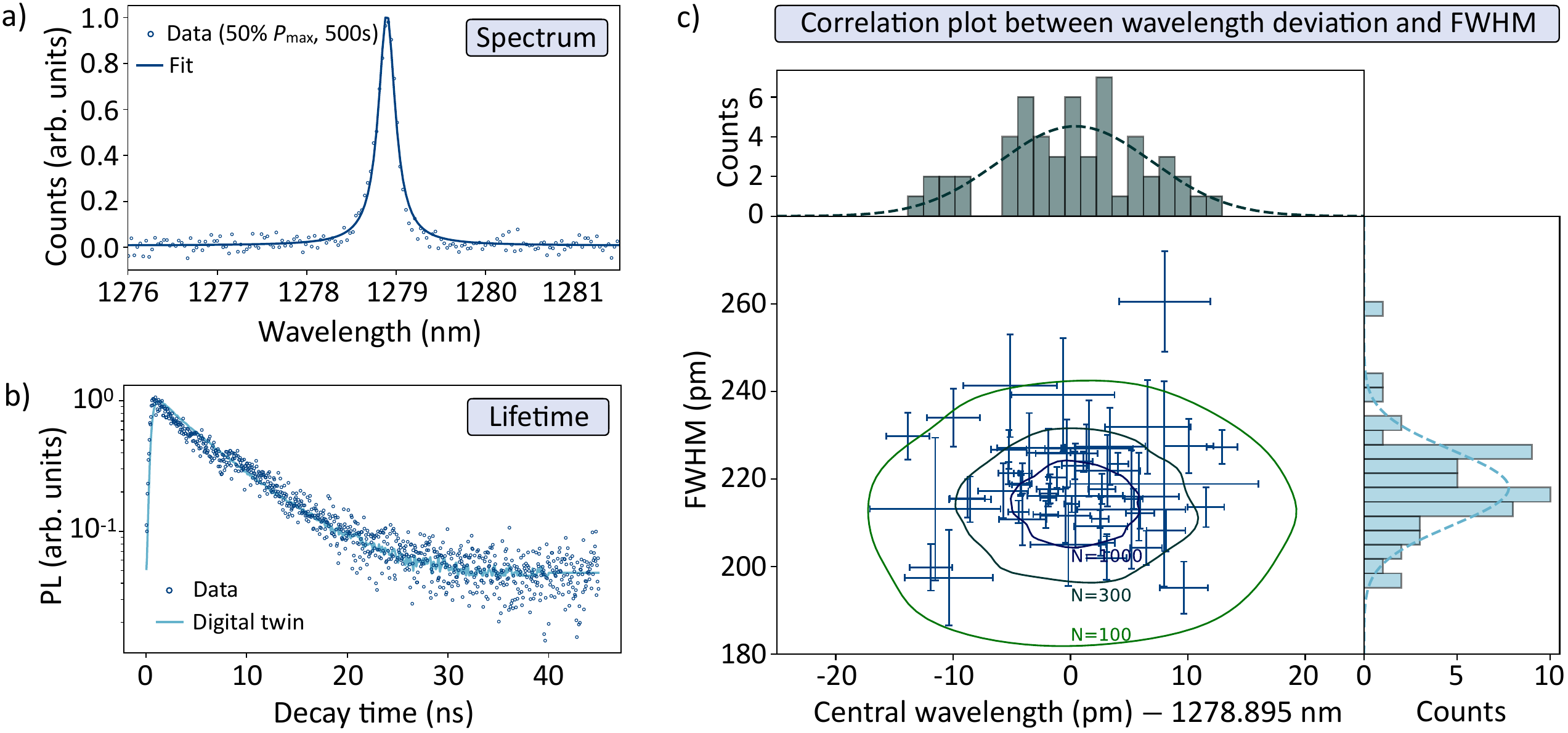}}
\caption{\textbf{Spectral characterization of laser-annealed emitters.} a) A representative spectrum of an annealed ensemble of emitters fitted with a Lorentzian function. b) Its corresponding lifetime data and fit modeled by our digital twin. c) Scatter plot of the central wavelength deviation from the nominal one of $1278.895~\mathrm{nm}$ and their corresponding full width half maxima (FWHM) for all ensembles measured across different annealed spots. The contours are the kernel density estimates of the digital twin-generated samples. The enclosed points represent the 95\% of total samples for a fixed inhomogeneous broadening of $88~\mathrm{pm}$ and different emitter numbers $N$. These contours indicate an estimated emitter count in the range $100-300$.} 
\label{fig3}
\end{figure*}

We now analyze the spectral properties of the generated emitters by varying annealing durations and powers, obtaining a PL grid of 72 annealed spots (reported in full in SI, Fig.~\ref{sfig:completegridannealing}). Fig.~\ref{fig3}a shows the spectrum of a representative activated ensemble, with a central wavelength of $(1278.896\pm0.002)~\mathrm{nm}$ and full width half maximum (FWHM) of $(0.227\pm0.005)~\mathrm{nm}$ --- consistent with the G-center ZPL --- extracted via Lorentzian fitting. Spectra were collected with an infrared spectrometer (see SI, Sec.~\ref{sec:setup}). Excited state lifetime measurements were performed using a pulsed $532~\mathrm{nm}$ laser along the excitation path. The results are shown in Fig.~\ref{fig3}b, together with the fit based on the transient-PL digital twin (see SI, Sec.~\ref{digitaltwinSI2}). When modeling the PL, the data-driven approach relies on fitting phenomenological models with parameters whose origin may not be strictly physical. For example, a single exponential decay is typically applied to the transient PL data though this is generally only valid for two-level systems excited with an idealized infinitesimally short optical pulse. This results in parameters that cannot be, at least in principle, corroborated separately by other physical measurements. In our approach, we use parameters that can be separately extracted from other measurements, such as the pulse shape, the surface recombination rate or the true atomic lifetime. In our system, both the surface recombination lifetime ($\tau_R$) and the atomic radiative lifetime ($\tau_a$) can give rise to the apparent exponential decay in the transient PL. In fact, by comparing the experimental data with the parametrized digital twin (see SI, Sec.~\ref{digitaltwinSI2}), we cannot determine unambiguously which parameter is the dominant factor. Both $\tau_a=(5.84\pm0.04)~\mathrm{ns}$, $\tau_R < 1~\mathrm{ns}$ and $\tau_R=(5.84\pm0.04)~\mathrm{ns}$, $\tau_a < 1~\mathrm{ns}$ can approximate the experimental data. The former combination is in agreement with what has been reported for both ensembles and single G-centers~\cite{lefaucher2023cavity, saggio2024cavity}, while the latter combination is consistent with the measurements of surface recombination rates for an SOI wafer~\cite{park2010influence}.

Moreover, we consider 54 spots annealed under different conditions to analyze the possible correlations between central wavelength deviation (from a nominal one of $1278.895~\mathrm{nm}$) and FWHM of each created ensemble. This is reported as a scatter plot in Fig.~\ref{fig3}c, and no correlation is observed. The peak wavelength distribution is very narrow (with a standard deviation of $6.2~\mathrm{pm}$ around the nominal wavelength). Notably, from interpolation of our digital twin data, we obtain that all our activated ensembles also present a very narrow inhomogeneous broadening of $\sim~88~\mathrm{pm}$ (see SI Sec.~\ref{statisticalAnalysisOfSpectraSI}), even under very different annealing conditions. 
The contours reported in Fig.~\ref{fig3}c are extracted by estimating the probability density function of the digital twin-generated samples. The enclosed points are 95\% of total samples for an inhomogeneous broadening of $88~\mathrm{pm}$ and different numbers of emitters $N$. Our digital twin allows us to further extract information on the emitters' numbers, with $N$ in the range $100-300$ (see SI Sec.~\ref{statisticalAnalysisOfSpectraSI}). 

\begin{figure}
  {\centering
  \includegraphics[width=\linewidth]{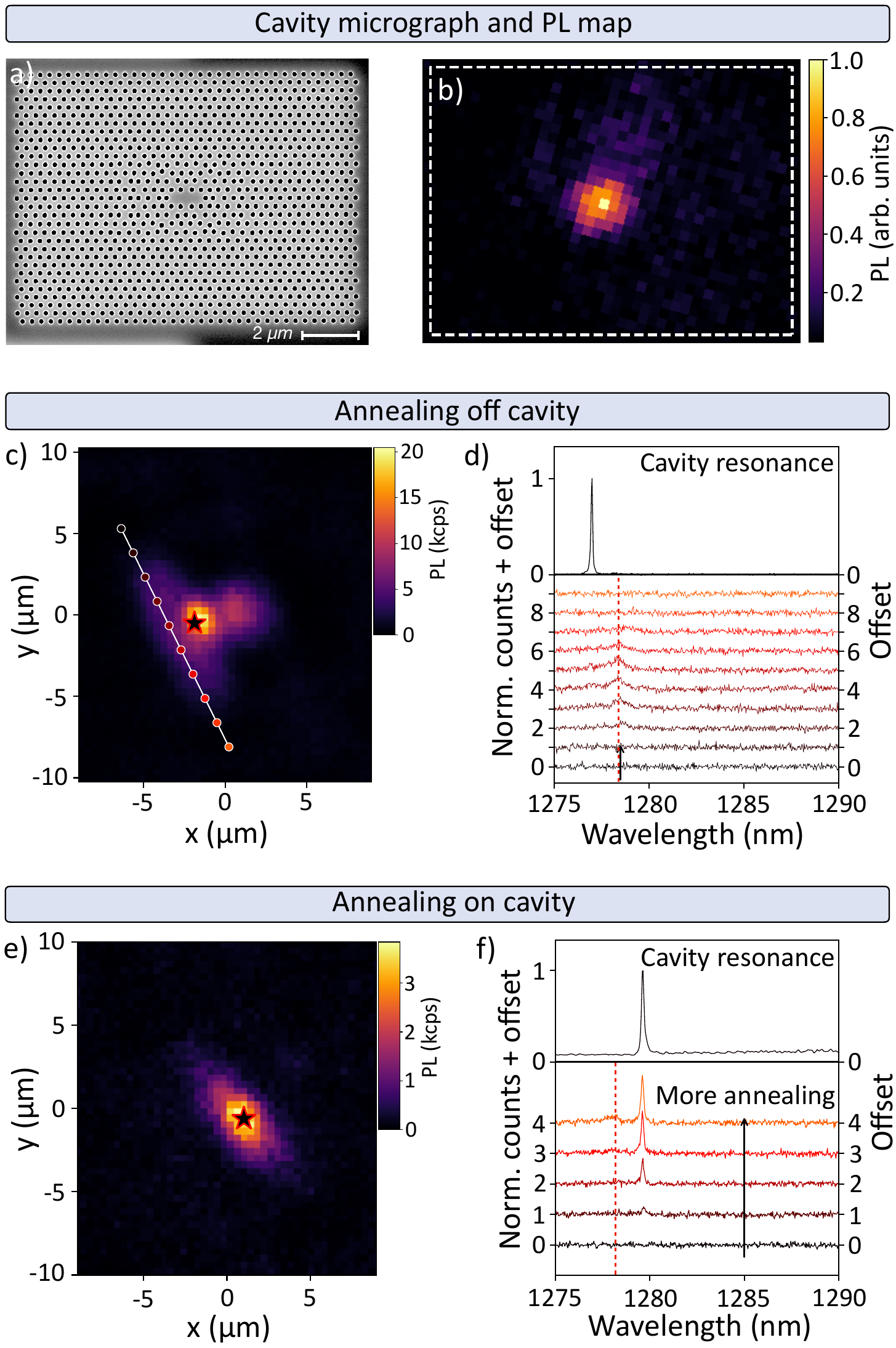}}
\caption{\textbf{Emitter formation off and on optical cavities.} a) Scanning electron micrograph of one of our photonic crystal cavities and b) a corresponding PL scan highlighting the PL emission from the cavity center. c) PL map of the cavity area showing the spots annealed at $4.8~\mathrm{mW}$ for $20~\mathrm{s}$ along the white line (i.e. on the photonic crystal structure). The cavity position is marked by the black star. d) PL measured on the cavity (black trace) along with spectra recorded at each annealing spot. Each spectrum is plotted in the same color as its corresponding dot in c). Emitter formation occurs on the photonic crystal structure and is highlighted by the red dotted line. e) PL map of the cavity area where repeated annealing rounds are performed on the cavity (star symbol). f) Experimental spectra and cavity resonance profile (black trace). In this case, as more annealing rounds are performed, peaks are observed that align to the cavity profile, along with smaller off-resonance peaks highlighted by the red dotted line.}
\label{fig4}
\end{figure}

Finally, we implement our annealing technique directly on pre-selected photonic crystal cavities optimized via inverse design~\cite{saggio2024cavity,panuski2022full}. 
A micrograph of a representative cavity and its corresponding PL map are shown in Figs.~\ref{fig4}a and b. These cavities are locally suspended to remove the underlying silicon dioxide, enabling emitter activation at $4.8~\mathrm{mW}$ due to heat localization --- much lower than the $\sim~30~\mathrm{mW}$ required for unpatterned SOI. Figs.~\ref{fig4}c-f show the difference between activating emitters on nanopatterned structures (photonic crystals) off the cavity and in the cavity. As illustrated in Fig.~\ref{fig4}c, laser annealing was performed on the photonic crystal structure on each indicated spot along the solid white line, avoiding the cavity area (black star). The spectra recorded for each spot are shown in Fig.~\ref{fig4}d and are color-matched to their corresponding dots in Fig.~\ref{fig4}c). Peaks near $1278.3~\mathrm{nm}$ (red dotted line) confirm emitter formation close to the cavity. Fig.~\ref{fig4}d also displays the PL measured on the cavity as the top black trace. As the emitters are being activated off the cavity, they do not align with the cavity resonance. Figs.~\ref{fig4}e and f show annealing performed directly on the cavity (star symbol in Fig.~\ref{fig4}e), leading to emitter creation within the cavity over time. The cavity resonance profile is the top black trace in Fig.~\ref{fig4}f. The more pronounced peaks are likely cavity-enhanced phonon sidebands of the created emitters, whose ZPLs at $1278.3~\mathrm{nm}$ (along the red dotted line) start emerging as additional annealing rounds are performed. These results demonstrate successful spatial alignment of color centers within optical cavities. 
Also in this case, we developed a digital twin of the laser annealing technique, now incorporating optical cavity parameters. In SI, Sec.~\ref{DTcavity}, we provide details on emitter formation in optical cavities along with discussions about the digital twin model. We demonstrate that our model successfully reproduces the experimental results for all cases shown in Fig.~\ref{fig4}.

\section{Discussion}
Previous works on laser annealing in diamond~\cite{ali2022laser} and silicon~\cite{jhuria2024programmable, andrini2024activation} use ultrafast pulsed laser sources. However, this comes at a cost not only in complexity, but also in the ablation of surfaces when applied to shallow targets. The threshold for surface ablation is often much lower than for emitter creation, and such ablation is attributed as the lead to degradation of optical coherence \cite{yurgens2021low} if not visible destruction of devices. Our method of CW thermal activation uses simpler hardware and is compatible with writing into the shallow nanophotonic structures that can power future quantum networks.

The emitters generated from unpatterned SOI did not show significant antibunching.  We attribute this to the limited collection efficiency of the current imaging system for detecting single emitters. With the ability to monitor the presence of single emitters, for example using photonic waveguides to enhance the collection efficiency~\cite{prabhu2023individually}, laser annealing is compatible with in-situ cryogenic measurements for closed-loop feedback between emitter characterization and controlled generation or erasure. With control on the annealing conditions, a single emitter can in principle be generated on demand starting from an unannealed sample, or by erasing from G-center ensembles. 

An exciting direction opened by this work is the possibility of generating color centers with very narrow inhomogeneous distributions in a targeted, on‐demand manner directly in cryogenic environments, using local thermal annealing in silicon or other semiconductor hosts. This capability would enable in-situ placement of single emitters within complex photonic~\cite{prabhu2023individually,redjem_single_2020} or electronic architectures~\cite{day2024electrical, dobinson2025electrically}, alleviating the need for blanket high‐temperature treatment or intricate ion‐implantation mask steps. Such deterministic “writing” of defect qubits could accelerate progress in large‐scale quantum networks, where multiple emitters must be precisely aligned to optical elements for efficient spin–photon interfaces both spatially and spectrally. Extending the CW annealing technique and the digital twin model to other host materials, such as diamond or silicon carbide, would broaden the palette of available color centers, potentially yielding defect species with desirable wavelengths or spin properties for quantum computation and sensing. 
Ultimately, it is desirable to have atom-cavity systems in the high-cooperativity regime. In-situ annealing provides the solution for achieving high coupling strength. In addition, the wafer-scale fabrication of photonic crystal cavities makes it possible to achieve quality factors of up to $Q\sim10^6$ \cite{panuski2022full}. Atom-cavity coupling is possible with several tuning approaches of either the emitters' ZPL --- e.g. tuning via electric fields~\cite{anderson_electrical_2019-2}, mechanical strain~\cite{wan_large-scale_2020-2}, or optical tuning~\cite{prabhu2023individually} --- or the cavity resonance --- e.g. via local thermal oxidation of silicon~\cite{panuski2022full}, gas tuning~\cite{li2015coherent,saggio2024cavity}, or temperature tuning~\cite{saggio2024cavity}. 
The precise control would become possible leveraging on recent reinforcement learning-based controllers~\cite{reuer2023realizing} that benefit from a first-principles simulator of the physical system, provided by the digital twin~\cite{li2024dynamic}.

\section{Conclusion}
We showed a reconfigurable activation of telecom emitters both in bulk and silicon photonic devices modeled by an end-to-end digital twin. Our digital twin unifies all ad-hoc phenomenological models into one first-principles model, able to dictate the right annealing conditions and predict the generated amount of PL and the physical system parameters. This work enables the targeted, deterministic generation of color centers with ultra-narrow inhomogeneous broadening and allows for precise emitter placement in photonic and electronic architectures. Our approach could advance large-scale quantum networks by ensuring precise spatial and spectral alignment for efficient spin–photon interfaces. 

\section*{Methods}
\subsection{Sample fabrication}
Following Ref. ~\cite{redjem_single_2020}, we start the fabrication process from a commercial SOI wafer with a $220~\mathrm{nm}$ silicon layer on a $2$~\textmu$\mathrm{m}$ silicon dioxide. We first implanted cleaved chips from this wafer with $^{12}$C with a dose of $5\times10^{13}~\mathrm{ions/cm}^{2}$ and energy of $36~\mathrm{keV}$, and then annealed at $1000~^{\circ}\mathrm{C}$ for $20~\mathrm{s}$. The samples were subsequently processed by a foundry (Applied Nanotools) for electron beam patterning and etching. The foundry also deposited a $2$~\textmu$\mathrm{m}$ silicon dioxide cladding using plasma-enhanced chemical vapor deposition (PECVD) at $300~^{\circ}\mathrm{C}$. To release the structures, we etched the cladding for $9~\mathrm{min}$ and $40~\mathrm{s}$ in a $15 \%$ hydrofluoric acid solution, followed by etching the buried oxide for $60~\mathrm{s}$ in a $49 \%$ hydrofluoric acid solution. Finally, the sample was dried with a critical point dryer.

\section*{Acknowledgements}
The authors acknowledge Dr Laiyi Weng, Dr Xingrui Cheng, Dr Jawaher Almutlaq, Dr Yong Hu for helpful discussions. V.S. acknowledges support from the Air Force Office of Scientific Research (AFOSR) under Award No. GR108261. C.Papon and A.B. acknowledge support from the NSF Convergence Accelerator program (Award No. 2040695) and from the NSF Engineering Research Center for Quantum Networks (Cooperative Agreement No. 1941583). C.E-H. acknowledges funding from the European Union’s Horizon 2020 research and innovation program under the Marie Sklodowska-Curie grant agreements No.896401, and from the Dutch Research Council (NWO, Project No. 601.QT.001). 
D.E. acknowledges support from the NSF RAISE TAQS program.
This material is based on research sponsored by the Air Force Research Laboratory (AFRL), under agreement number FA8750-20-2-1007. 
The U.S. Government is authorized to reproduce and distribute reprints for Governmental purposes notwithstanding any copyright notation thereon.
The views and conclusions contained herein are those of the authors and should not be interpreted as necessarily representing the official policies or endorsements, either expressed or implied, of the Air Force Research Laboratory (AFRL), or the U.S. Government.

\bibliography{Si_artificial_atoms.bib}

\resumetoc

\newpage
\onecolumngrid
\appendix
\renewcommand{\thefigure}{S\arabic{figure}}
\setcounter{figure}{0}
\newpage 

\section*{Supplementary Information:\\ End-to-end physics-based modeling of laser-activated color centers in silicon}

\begin{adjustwidth}{80pt}{80pt}
\centering
Qiushi Gu$^{1}$, Valeria Saggio$^{1}$, Camille Papon$^{1}$, Alessandro Buzzi$^{1}$, Ian Christen$^{1}$,\\ Christopher Panuski$^{1}$, Carlos Errando-Herranz$^{1,2,3}$, and Dirk Englund$^{1}$
\end{adjustwidth}

\vspace{0.07cm}

\begin{adjustwidth}{80pt}{80pt}
\centering
\small{\textit{$^1$Massachusetts Institute of Technology, Cambridge, MA, USA\\
$^2$QuTech and Kavli Institute of Nanoscience, Delft University of Technology, Delft, Netherlands\\
$^3$Department of Quantum and Computer Engineering, Delft University of Technology, Delft, Netherlands}} 
\end{adjustwidth}

\setstretch{2}
\tableofcontents
\setstretch{1}
\newpage 

\subsection{Experimental setup}
\label{sec:setup}
\begin{figure}[b]
  {\centering
  \includegraphics[width=0.88\textwidth]{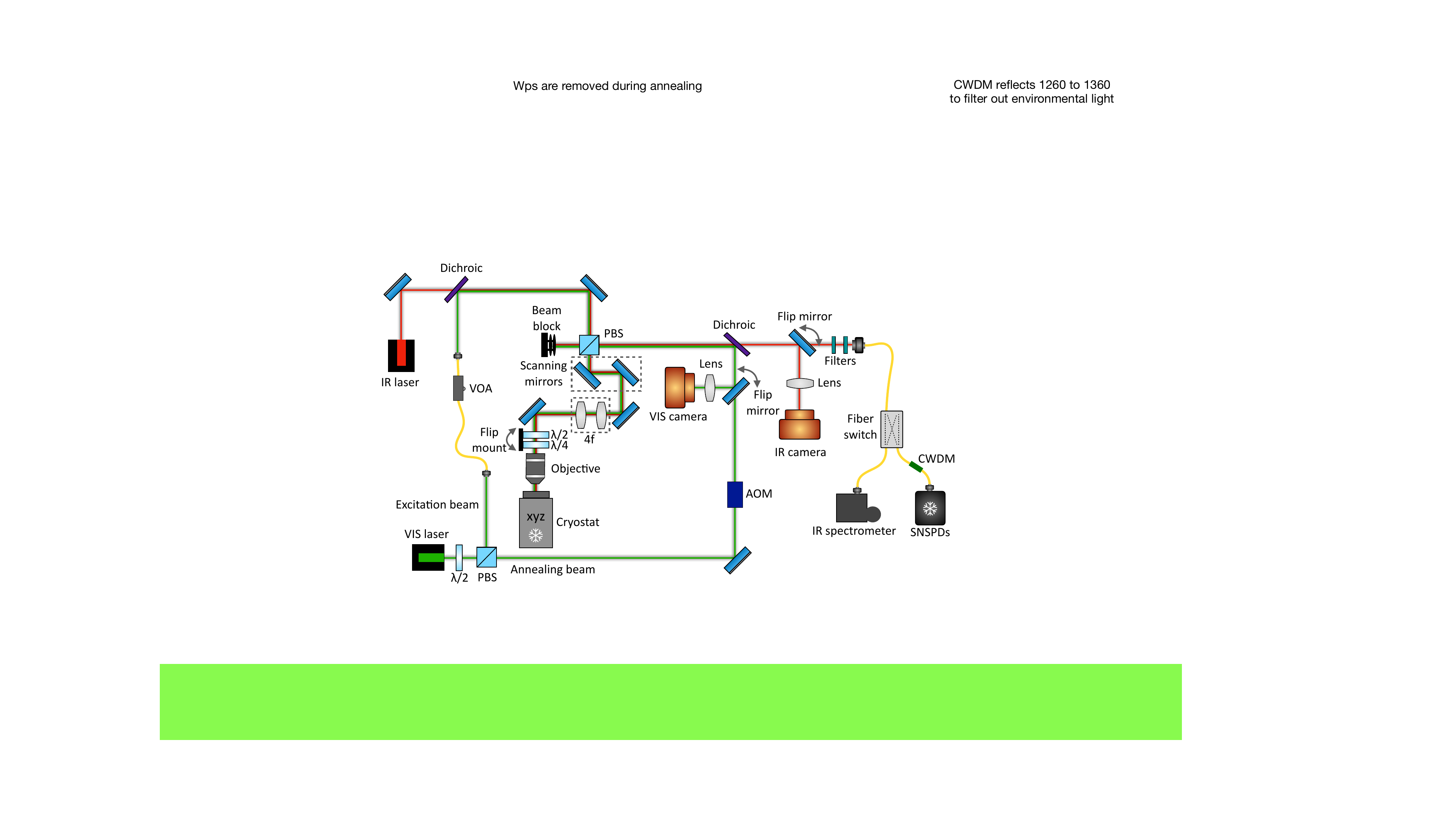}}
\caption{\textbf{Setup schematic.} IR laser light and visible laser light (excitation beam) shine through a PBS, a set of scanning mirrors, a 4f system, and polarization rotation components, into an objective and into the cryostat, where the sample is placed. The IR light reflected off the sample, along with the IR photon emission, travel backwards along the same incident path and are reflected off the PBS, going to an IR camera for imaging or to a spectrometer or SNSPDs for detection, respectively. The reflected excitation beam is sent to a visible camera to enable precise beam alignment. A second visible laser beam (annealing beam) is directed onto the sample using a dichroic mirror. Our filtering station includes two filters --- one long pass and one short pass --- along with a CWDM. A VOA is used on the excitation laser path, and an AOM on the annealing laser path for power control.}
\label{sfig:m3setup}
\end{figure}

Our measurement setup is illustrated in Fig.~~\ref{sfig:m3setup}. A continuous-wave (CW) visible (VIS) laser beam is split at a polarizing beam splitter (PBS) after passing through a half-wave plate ($\lambda/2$) for power control. The reflected component (excitation beam), is sent to a dichroic mirror after passing through a variable optical attenuator (VOA) enabling automated power variations. The dichroic mirror combines the VIS light with infrared (IR) laser light. The two beams are then transmitted through a polarizing beam splitter (PBS), and illuminate the sample --- placed in the cryostat --- after passing through two scanning mirrors, a 4f system composed of two lenses, a set of a $\lambda/2$ and a quarter-wave plate ($\lambda/4$), and an objective. Their reflections from the sample, together with the IR photon emission coming from excitation with the VIS excitation beam, follow the same path backwards up to the PBS, where they get reflected and then separated at another dichroic mirror. The VIS beam is sent to a visible camera to enable precise alignment with the IR laser beam. The IR light reflected from the surface of the sample can be sent to an IR camera for imaging, and the IR photon emission to either an IR spectrometer, or to our superconducting nanowires single-photon detectors (SNSPDs) with the use of a fiber switch. Two filters --- one longpass with cutoff wavelength at $1250~\mathrm{nm}$ and one shortpass with cutoff wavelength at $1300~\mathrm{nm}$ --- are used to filter a $50~\mathrm{nm}$-wide region around the emitters' emission wavelength. A coarse wavelength division multiplexing (CWDM) is used to filter out the environmental light and thus reduce the SNSPDs' dark counts. In order to perform laser annealing, the transmitted component of the VIS laser from the first PBS (annealing beam) is used. It is sent through an acousto-optic modulator (AOM) for power modulation before being routed onto the sample through a dichroic mirror. Mirrors mounted on flip mounts allow us to switch between annealing or visible imaging in the case of the VIS laser light, or between IR imaging or photon collection in the case of the IR light. 

The VIS light is used for exciting or annealing the emitters, while the IR laser is used for both cavity characterization and imaging of our sample. The cavities are characterized in cross-polarization, as detailed in our previous work~\cite{saggio2024cavity}. For this measurement, a $\lambda/2$ and a $\lambda/4$ are needed before the objective. These wave plates are mounted on a flip mount, and are flipped away during the annealing process. 

For annealing, exciting and characterizing our emitters we use a CW Coherent Verdi G5 at $532~\mathrm{nm}$. For spectral characterization, we use an IR spectrometer consisting of a PyLon IR CCD from Princeton Instruments and a grating with a density of $900~\mathrm{gr/mm}$ and a $1.3$~\textmu$\mathrm{m}$ blaze, leading to a pixel-defined resolution of $40~\mathrm{pm}$. For lifetime measurements, we use a pulsed laser from NKT Photonics (SuperK) following the excitation path. This laser, not shown in the figure, features a maximum repetition rate of $78~\mathrm{MHz}$ and is filtered by a bandpass filter centered at $532~\mathrm{nm}$. Our IR laser is a superluminescent diode S5FC1018S from Thorlabs operating at $600~\mathrm{mA}$, with a broadband emission centered at $1310~\mathrm{nm}$. Our objective is a collar-corrected objective LCPLN50XIR from Olympus with a numerical aperture (NA) of $0.65$, and is mounted externally to the cryostat. Our cryostat is a Montana Instruments system, operating at a temperature of $\sim 7~\mathrm{K}$. The sample is mounted on a XYZ cryogenic piezoelectric stage from Attocube. Our SNSPDs from Photon Spot feature detection efficiencies of around $20\%$, and are readout with a Swabian Instruments Timetagger~20. The visible camera is a Thorlabs Zelux, and the IR camera is an InGaAs cooled CCD camera from Allied Vision Goldeye.

\subsection{First-principles digital twin of the annealing process}
\label{digitaltwinSI}
In this section, we elaborate on the development of the digital twin for our annealing technique. Our digital twin consists of two parts. The first one, outlined in the following subsection~\ref{digitaltwinSI1}, considers atomic diffusion processes to model the spatial profile of the generated PL. The second one, described in subsection~\ref{digitaltwinSI2}, takes into account atomic rates to model transient PL measurements.

\subsubsection{Thermal transport and atomic diffusion}
\label{digitaltwinSI1}

\begin{table}[b]
    \centering
        \begin{tabular}{|c|l|c|}
        \hline
        Parameter & Value used in digital twin & Comment \\ \hline
        $k_f^0$ & $5.4\times10^{9}~\mathrm{m^3\,s^{-1}\,mol^{-1}}$ & $21.8\times10^{9}$~\cite{brenet2015atomistic}, $358\times10^{9}$~\cite{davies1988metastable} \\ \hline
        $E_a^f$ & $(0.826\pm0.01)~\mathrm{eV}$ & $0.938~\mathrm{eV}$~\cite{davies1988metastable}, $0.58~\mathrm{eV}$~\cite{brenet2015atomistic}\\ \hline
        $k_b^0$ & $1\times 10^{12}~\mathrm{s}^{-1}$ &  Taken from~\cite{brenet2015atomistic}\\ \hline
        $E_b^f$ & $(1.962\pm0.01)~\mathrm{eV}$ & $(1.70\pm0.05)~\mathrm{eV}$~\cite{davies1989annealing},$1.86~\mathrm{eV}$~\cite{brenet2015atomistic} \\ \hline
        
        $k_2^0$ & $20~\mathrm{m^3\,s^{-1}\,mol^{-1}}$  &  Taken from~\cite{davies1989annealing}\\ \hline
        $E_a^2$ & $0.4~\mathrm{eV}$ & Taken from~\cite{brenet2015atomistic}\\ \hline
        $k_{-2}^0$ & $3\,\mathrm{m^3\,s^{-1}\,mol^{-1}}$ & Taken from~\cite{davies1989annealing} \\ \hline
        $E_a^{-2}$ & $1.85\,\mathrm{eV}$ &  Taken from~\cite{brenet2015atomistic}\\ \hline
        \end{tabular}
            \caption{Parameters used in the digital twin based on atomic diffusion. }
    \label{tab:dt-comsol}
\end{table}

At the atomic scale, a sequence of complex atomic diffusion processes is believed to happen. To varying degrees of complexity, mechanisms have been proposed to account for activation of dynamical thermal responses in earlier spectroscopic studies. We start from the following meta-reactions,
\begin{align}
\ce{C_i + C_s &<=>[{\textit{k}_f}][{\textit{k}_{b}}] C_iC_s} \label{eq-dt-comsol1},\\
\ce{C_s + I &<=>[{\textit{k}_2}][{\textit{k}_{-2}}] C_i}, \label{eq-dt-comsol2}
\end{align}
where \ce{C_i}, \ce{C_s} and \ce{I} are interstitial carbon, substitutional carbon and interstitial silicon atoms respectively. $k_{\text{f}}$ and $k_{\text{b}}$ are the forward and backward rate constants detailed later in this section, while $k_2$ and $k_{-2}$ are the forward and backward secondary rate constants. Both reactions are reversible, such that the forward and backward elementary reactions are in constant thermal equilibrium. Each elementary reaction step follows Arrhenius law, for example,
\begin{align}
    \frac{d}{dt}[\ce{C_i}] &= k_{b}[\ce{C_iC_s}], \\
    \frac{d}{dt}[\ce{C_iC_s}] &= k_{f}[\ce{C_i}][\ce{C_s}],
\end{align}
where $[\cdot]$ is used to indicate the atomic concentration, $k_{b}=k_{b}^0\exp\left(-\frac{E_a^{b}}{k_\mathrm{B}T}\right)$ and $k_{f}=k_{f}^0\exp\left(-\frac{E_a^{f}}{k_\mathrm{B}T}\right)$ with $k_{\mathrm{B}}$ Boltzmann constant and $T$ the absolute temperature. The constants $k_{b}^0$, $k_{f}^0$ are Arrhenius prefactors and $E_a^{b}$, $E_a^{f}$ are the activation energies determined from the experimental data. The prefactors turn out to be less important since a small error on the activation energy significantly changes the rate constants due to the exponentiation. We also extract $k_{b}^0$ from the experimental data. The series of parameters used in our model is detailed in Table~\ref{tab:dt-comsol}.

Eq.~\ref{eq-dt-comsol1} describes the reaction where the combination of interstitial and substitutional carbon atoms forms G-centers (\ce{C_iC_s}) and the reversible dissociation of G-centers forms individual carbon atoms. Moreover, a substitutional carbon atom and an interstitial silicon atom can swap positions via the Watkins replacement reaction given in Eq.~\ref{eq-dt-comsol2}, which gives rise to constant thermal hopping from site to site. 

The initial concentrations were taken from past literature where it is known that ion implantation primarily produces interstitial carbon~\cite{zhiyenbayev2023scalable} and interstitial silicon atoms with similar concentrations (\ce{[C_i]}$=300\times 10^{16}~\mathrm{atoms\,cm^{-3}}$ and \ce{[I]}$=300\times 10^{16}~\mathrm{atoms\,cm^{-3}}$). The initial concentrations of both \ce{[C_iC_s]} and \ce{[C_s]} are of much smaller values, with \ce{[C_iC_s]}$=0.74\times 10^{16}~\mathrm{atoms\,cm^{-3}}$ and \ce{[C_s]}$=0.1\times 10^{16}~\mathrm{atoms\,cm^{-3}}$. The activation and dissociation energies of the Watkins replacement reaction are directly taken from Ref.~\cite{brenet2015atomistic}. Each reaction has a temperature-dependent rate and this is coupled to the thermal transport model at each point in space. We use COMSOL multiphysics to simulate the coupled heat transport and atomic reactions. For thermal transport, we assume that a Gaussian beam of diameter $D$ is incident on a silicon-on-insulator (SOI) wafer with $220~\mathrm{nm}$ of device layer, $2$~\textmu$\mathrm{m}$ of silicon dioxide layer and a $700$~\textmu$\mathrm{m}$ handle layer. The bottom surface is set to the cryostation temperature of $\sim 10~\mathrm{K}$, while the top and side surfaces are thermally insulated due to surrounding vacuum. The laser beam size $D=1.25$~\textmu$\mathrm{m}$ is separately determined from the wide field image of the sample surface. Furthermore, we measured the point spread function size of the imaging system to be approximately $1$~\textmu $\mathrm{m}$ using a wide field camera. The laser power is directly calibrated after the cryostation window to accurately reflect the incident annealing laser power on the sample. Using this physical setting, we compute the density of G-centers as a function of space, annealing duration and annealing laser power using the FEM solver, COMSOL Multiphysics.

From the FEM simulation, we obtain the emitter density $\sigma(x,y)$ for a particular combination of annealing duration and laser power. Next, we generate the confocal PL map, $I(X, Y)$, from this distribution. We spatially discretize the emitter density into grids of $300~\mathrm{nm}$ (arbitrarily chosen) in size and compute the number of emitters per grid, $n(x_i, y_i)$ using a Poissonian random number generator. For each pixel on the PL map, $(X_j, Y_j)$, we assume a green excitation laser centered around $(X, Y)$ and each grid of emitters experiences $P_{ex}(x_i, y_i|X_j, Y_j)$ excitation power, which is a Gaussian profile centered around $(X_j, Y_j)$ and measured at $(x_i, y_i)$. These emitters then contribute an emission of $P_{em}(X_j, Y_j|x_i, y_i)$, which is another Gaussian profile centered around $(x_i, y_i)$ and measured at $(X_j, Y_j)$. The final counts on pixel $(X_j, Y_j)$ are estimated by summing over all emitter grids $(x_i, y_i)$. This is repeated for each pixel $(X_j, Y_j)$ to generate the confocal map $I(X_j, Y_j)$.

This model takes in the experimental parameters of annealing duration and power and generates a confocal map that can be compared directly with experiments. We assume no free parameters, except the activation energies which we fit from the experimental data. These values are close to the literature values. 

\subsubsection{Master equation modeling of the electron capture and transient PL measurements}
\label{digitaltwinSI2}

In this section, we describe the digital twin involving atomic transition rates to generate transient PL measurements. 

\begin{figure}[t]
  {\centering
\includegraphics[width=7cm]{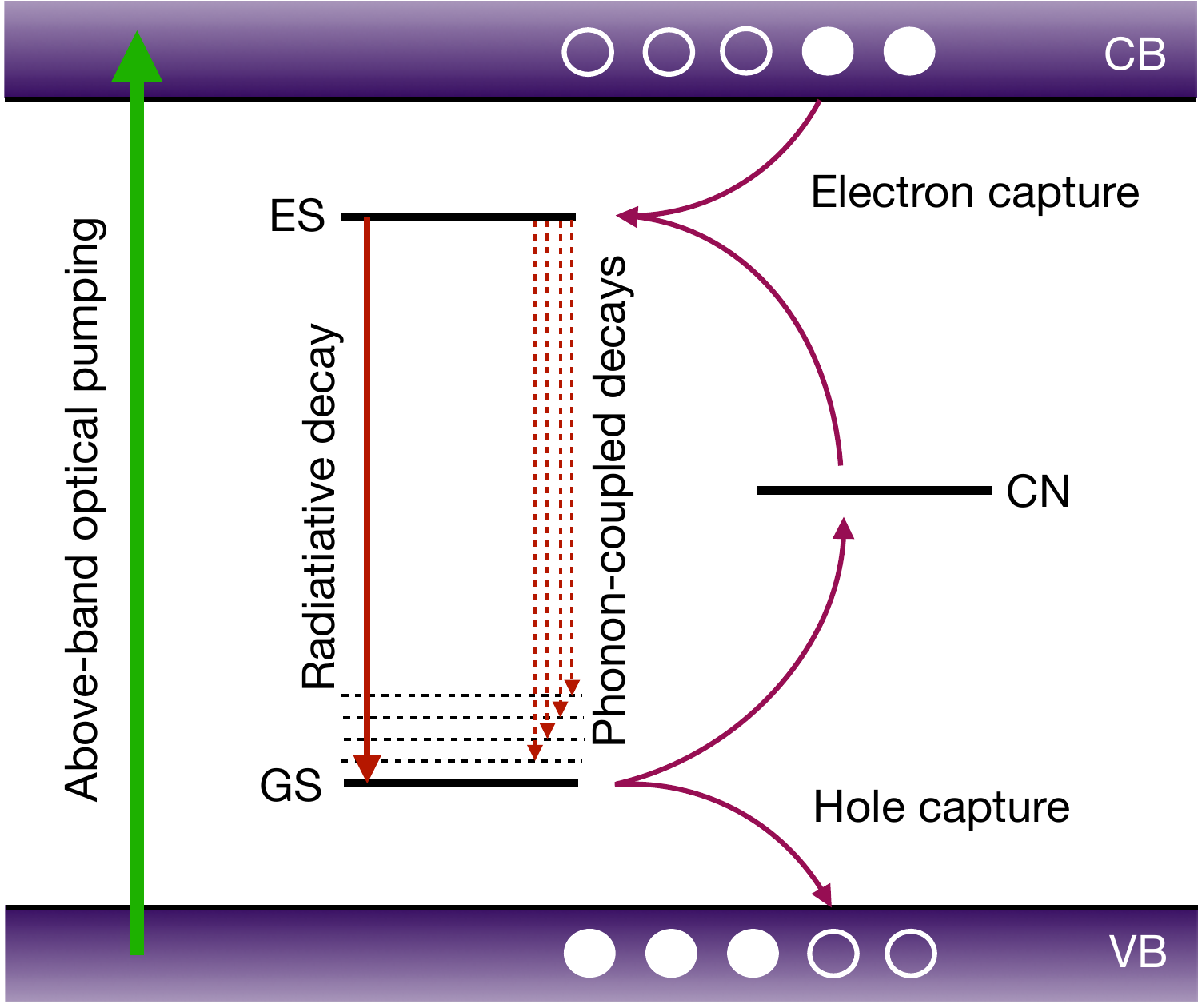}}
\caption{\textbf{Illustration of energy levels.} Above-band optical pumping (green arrow) from the valence band (VB) to the conduction band (CB) generates local electrons and holes (filled and empty circles, respectively), which quickly diffuse toward the surface and the surrounding areas. A small fraction of these electrons is captured by local charge-neutral (CN) G-centers (i.e. G-centers that do not have an overall electric charge), exciting them to the excited state (ES). The excited G-center then radiatively decays to the ground state (GS) (solid red arrow), giving rise to the zero-phonon line (ZPL), or to phonon-coupled ground states (dashed red arrows), emitting photons into the phonon sideband. The GS subsequently releases the electrons to the VB and returns into the CN state.}
\label{figS-LifetimeDigitalTwin}
\end{figure}

The off-resonance optical pumping of G-centers is believed to start with free carrier generation and subsequent capture by local defects. Particular steps of this mechanism are still unclear, for example if and how many shelving states are present, and how the electron capture process takes place. We take the silicon antisite defect in silicon carbide near a stacking fault as a reference~\cite{khramtsov2018enhancing} and show that this model is also consistent with existing experimental data. We thus adopt a similar technique to model this from first principles by considering the incoherent evolution of the valence and the conduction bands, modeled as a two-level system, and the defect, modeled as a three level system as illustrated in Fig.~\ref{figS-LifetimeDigitalTwin}. There are four critical steps in this model, detailed in what follows.  

\begin{enumerate}
    \item The optical pump (green arrow in Fig.~\ref{figS-LifetimeDigitalTwin}) induces a local generation of charge carriers, meaning that electrons (filled white circles) in the valence band (VB) are promoted to the conduction band (CB), leaving holes (empty circles) in the VB;
    \item the defect (G-center in our case) captures local electrons with a probability proportional to their concentration, thus transitioning to the excited state (ES);
    \item the defect relaxes back to the ground state (GS) by emitting a photon. This transition can be phonon-assisted (dashed red arrows), resulting in photons being emitted into the phonon sideband, or occur without any phonon mediation (solid red arrow), resulting in photon emission into the zero-phonon line (ZPL);
    \item lastly, the defect captures a hole and becomes charge-neutral (CN) again. 
\end{enumerate}

In particular the first step is affected by local carrier recombination processes (in this case, surface-dominated recombination~\cite{park2010influence}).

To capture the change in conduction band and valence band populations (denoted by $\rho_{\mathrm{CB}}$ and $\rho_{\mathrm{VB}}$), we use the rate equation
\begin{equation}
\frac{d\rho_{\mathrm{CB}}}{dt} = G\rho_{\mathrm{VB}}-\Gamma_{r}\rho_{\mathrm{CB}},
\end{equation}
which takes into account both the incoherent above-band pumping process, with $G$ being the pumping rate, and the carrier recombination process, with $\Gamma_r$ being the recombination rate. Both $G$ and $\Gamma_r$ are taken as free parameters. Since the optical pumping is small compared to the absorption limit of silicon, we assume that $G$ is proportional to the optical power. The electron and hole capture rates are proportional to the local concentration of the respective carriers $c_n$ and $c_p$ and are $\Gamma_n = c_n \Gamma_0$ and $\Gamma_p = c_p \Gamma_0$ with $c_n=c_p$. The parameter $\Gamma_0$ is related to the carrier capture cross-section and the diffusion rate for the respective carriers.

While there are local variations in carrier concentrations, electron capture by G-centers is still small. To model the dynamics after excitation of G-centers, we use the Lindbladian equation (or the Master equation)
\begin{equation}
    \frac{d\rho}{dt}=-\frac{i}{\hbar}\left[H, \rho\right] + \sum_k L_k\rho L_k^\dagger - \frac{1}{2} \left(L_k^\dagger L_k \rho+\rho L_k^\dagger L_k\right),
\end{equation}
where $\rho$ is the system density operator spanned by the ground state $\ket{\mathrm{GS}}$, excited state $\ket{\mathrm{ES}}$ and charge-neutral state $\ket{\mathrm{CN}}$ of the G-center as well as the phonon states coupled to the ground state, and $H$ is the Hamiltonian describing the system. There is discussion about a dark metastable triplet state, previous observed in emitter ensembles~\cite{lee_optical_1982, odonnell_origin_1983}. However as its absence still allows the digital twin to accurately model the experimentally observed transient PL, the metastable state was not included in the energy levels.

The incoherent evolutions are introduced via collapse operators $L_k$, where each $k$ corresponds to one decoherence channel to be described next. These operators incorporate the following terms into the Master equations: the radiative decay term $\sqrt{\xi_0\Gamma_a}|\mathrm{GS}\rangle\langle\mathrm{ES}|$ (with $\xi_0$ branching ratio into the ZPL and $\Gamma_a$ radiative decay rate of the excited state), dephasing term $\sqrt{\Gamma_2}\left(|\mathrm{ES}\rangle\langle\mathrm{ES}|-|\mathrm{GS}\rangle\langle\mathrm{GS}|\right)$ (with $\Gamma_2$ being the dephasing rate), the charge capture term $\sqrt{\Gamma_n}|\mathrm{ES}\rangle\langle\mathrm{CN}|$, and the hole capture term $\sqrt{\Gamma_p}|\mathrm{CN}\rangle\langle\mathrm{GS}|$. 
The dephasing rate is chosen to be $\Gamma_2=6.8~\mathrm{GHz}$ to align with the linewidth of single emitters' homogeneous broadening measured in Ref.~\cite{redjem2023all}. The optical pumping enters the Master equation through the time dependence of $\Gamma_{n,p}(t)=c_{n,p}(t)\Gamma_0$.

The coherent evolution is modeled by the Hamiltonian $H_0=\omega_a |\mathrm{ES}\rangle\langle\mathrm{ES}|$, where $\omega_a=2\pi \frac{c}{\lambda}$. $\lambda=1279~\mathrm{nm}$ is the ZPL of G-centers. 

Additionally, we introduce phononic coupling to account for the phonon sideband typically observed in experiments. The phononic coupling has a coherent part $H_{\mathrm{ph}}=n\omega_{\mathrm{ph}} |\mathrm{GS}\rangle|\mathrm{ph, n}\rangle\langle\mathrm{ph, n}|\langle\mathrm{GS}|$, where $|\mathrm{GS}\rangle|\mathrm{ph, n}\rangle$ is the ground state coupled to $n$ phonons and $\hbar \omega_{\mathrm{ph}}$ is the energy of each phonon, and an incoherent part  $\xi_n\Gamma_a |\mathrm{ES}\rangle\langle\mathrm{ph, n}|\langle\mathrm{GS}|$, where $n=1,2,\cdots,12$ is the $n^{\mathrm{th}}$ phononic level coupled to the ground state and is chosen to be 12 in our digital twin. The ratios $\xi_0, \xi_1, \cdots, \xi_{12}$ are the branching ratios indicating the probability of decaying through a particular pathway with $\sum_{i=0}^{12}\xi_i=1$, $\xi_0/\xi_1=50$ and $\xi_1=\xi_2=\cdots=\xi_{12}$. The phononic energy $\omega_{\mathrm{ph}}/2\pi$ is chosen to be $500~\mathrm{GHz}$. The phononic energy and the number of phonon modes are chosen so as to approximate the slowly varying phonon sideband which extends to $1300~\mathrm{nm}$, and the branching ratios are also chosen such that the generated spectrum aligns with single emitter measurements~\cite{prabhu2023individually, hollenbach_wafer-scale_2022}. It however does not reflect the measured emission into the ZPL (approximately 15\%~\cite{lefaucher2023cavity}) which requires accounting for the non-radiative decay into the shelving states not considered in our digital twin.

The computation is carried out using the Master equation solver in the Python QuTiP package, taking the charge carrier capture rate, surface recombination time, and radiative rate of the atomic system as input parameters, subject to a Gaussian pump pulse of fixed finite FWHM of $600\,\mathrm{ps}$ measured using a time tagger.

By solving the time-dependent Master equation, we obtain the time-dependent density operator $\rho(t)$, from which the transient PL is obtained as $\mathrm{PL}(t)\propto \Gamma_a \mathrm{Tr}\left(\rho(t)|\mathrm{ES}\rangle\langle\mathrm{ES}|\right)$. 

However, as solving the Master equation for various combinations of input parameters is computationally intensive, we generated a large number of combinations of the three input parameters ($\Gamma_r$, $\Gamma_a$, $\Gamma_0$) within a range of $10~\mathrm{MHz}$ to $5~\mathrm{GHz}$ with logarithmic normal distribution, and used a multilayer perceptron (MLP) --- a type of artificial neural network composed of multiple layers where each layer contains neurons~\cite{popescu2009multilayer} --- to approximate the Master equation solver. This enables parallelizing the solver. Using the MLP, we can approximate any input parameters within the range by assuming that the output varies smoothly with the input. The MLP is the first-principles digital twin in this case, and we compare the physical measurements and the digital twin to determine the best combination parameters. Using curve fit on this model, we deduce the rates reported in the main text.

\begin{figure}[t]
  {\centering
\includegraphics[width=\linewidth/2]{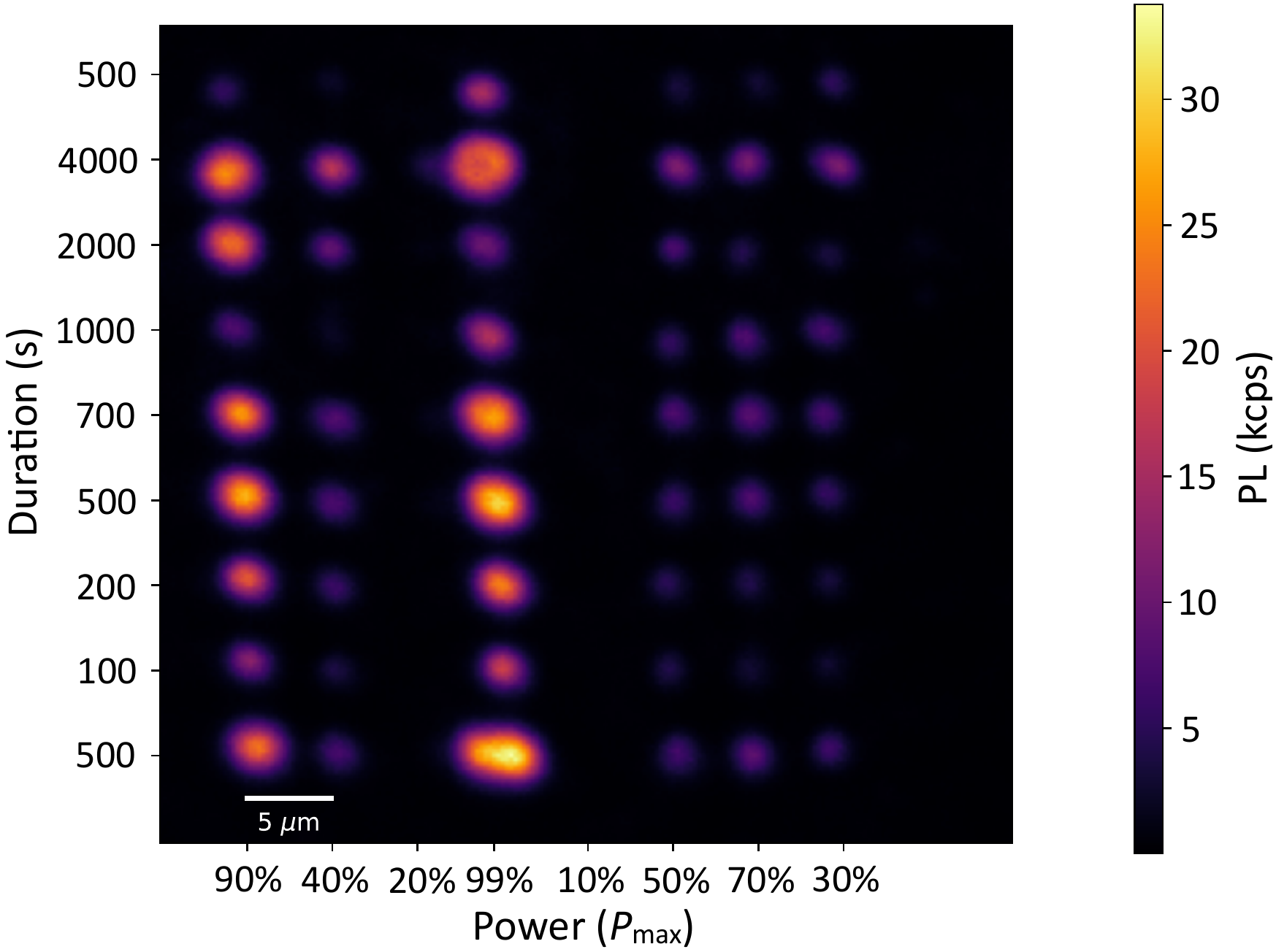}}
\caption{\textbf{PL grid for different annealing conditions.} The generated PL is shown for different annealing durations (vertical axis) and powers (horizontal axis). The laser power ranges from $10\%$ to $99\%$ of the maximum available power $P_{\mathrm{max}}=81~\mathrm{mW}$ at the sample.} 
\label{sfig:completegridannealing}
\end{figure}

\subsection{Different annealing conditions}
\label{differentannealing}
In addition to the PL map of annealed emitters reported in Fig.~\ref{fig1} in the main text, we show the complete grid of annealed emitters in Fig.~\ref{sfig:completegridannealing}, which includes more annealing conditions --- specifically higher percentages of annealing maximum power $P_{\mathrm{max}}$ up to $99\%$, with $P_{\mathrm{max}}=81~\mathrm{mW}$, and different annealing durations. There is variability in the spots' brightness potentially due to local variations of carbon density, surface termination, strain, or other material-related inhomogeneities. These measurements are used to carry out the spectral analysis presented in Fig.~\ref{fig3} in the main text.

\subsection{Statistical analysis of G-center ensembles}
\label{statisticalAnalysisOfSpectraSI}
To model the spectra of ensembles of emitters (measured with the IR spectrometer), we consider the ensemble spectrum to be formed by the sum of individual emitters' spectra. 
For a single emitter, the measured spectrum is a convolution of the gaussian instrument response function (IRF) --- limited by the spectrometer grating, with a 
width at $1/e$ of $\sigma_0=24~\mathrm{pm}$ --- with the intrinsic emitter Lorentzian lineshape 
($6.8~\mathrm{GHz}$, or approximately $\gamma_0=37~\mathrm{pm}$ in the wavelength domain). This convolution, indicated with $\mathrm{PL}_{\mathrm{s}}^i(\lambda)$ where $\mathrm{s}$ refers to single emitters and $i$ to the $i^{\mathrm{th}}$ emitter, results in a Voigt lineshape $V$ centered around $\lambda_0^i$ with background $B$ and amplitude $\mathrm{PL}_0$, leading to $\mathrm{PL}^i_{\mathrm{s}}(\lambda)=\mathrm{PL}_0 V(\lambda-\lambda_0^i; \sigma_0, \gamma_0)+B$. The ensemble spectrum consisting of $N$ emitters will therefore read
\begin{align*}
\mathrm{PL_{ens}}(\lambda) & =\sum_{i=1}^N\mathrm{PL}^i_{\mathrm{s}}(\lambda).
\end{align*}

Variations in the emitters' local environment lead to inhomogeneous broadening, which results in distinct spectral properties for each emitter and thus a broadened emission spectrum for the ensemble. Without \textit{a priori} knowledge, we assume that the central wavelengths of individual emitters follow a Gaussian distribution  $\lambda^i_{0{\mathrm{s}}}\sim\mathcal{N}(\lambda_0, {\Delta \lambda_{\mathrm{s}}}^2)$, where the standard deviation $\Delta \lambda$ is the inhomogeneous broadening.

For a large number of emitters $N$ ($N\gg\Delta \lambda/\sigma_0\sim 70$, extracted from the data) generated on the $j^{\mathrm{th}}$ annealed spot, the final ensemble lineshape is close to a Lorentzian lineshape with a central wavelength $\lambda_{0\mathrm{ens}}^j$ and FWHM $\Gamma_{\mathrm{ens}}$:
\begin{equation}
\mathrm{PL}_{\mathrm{ens}}^j(\lambda) = \mathrm{PL_0} \frac{1}{1+\left(\frac{\lambda-\lambda_{0ens}^j}{{\Gamma_{\mathrm{ens}}}/{2}}\right)^2}.
\end{equation}

\begin{figure}[t]
    \centering
    \includegraphics[width=0.72\linewidth]
    {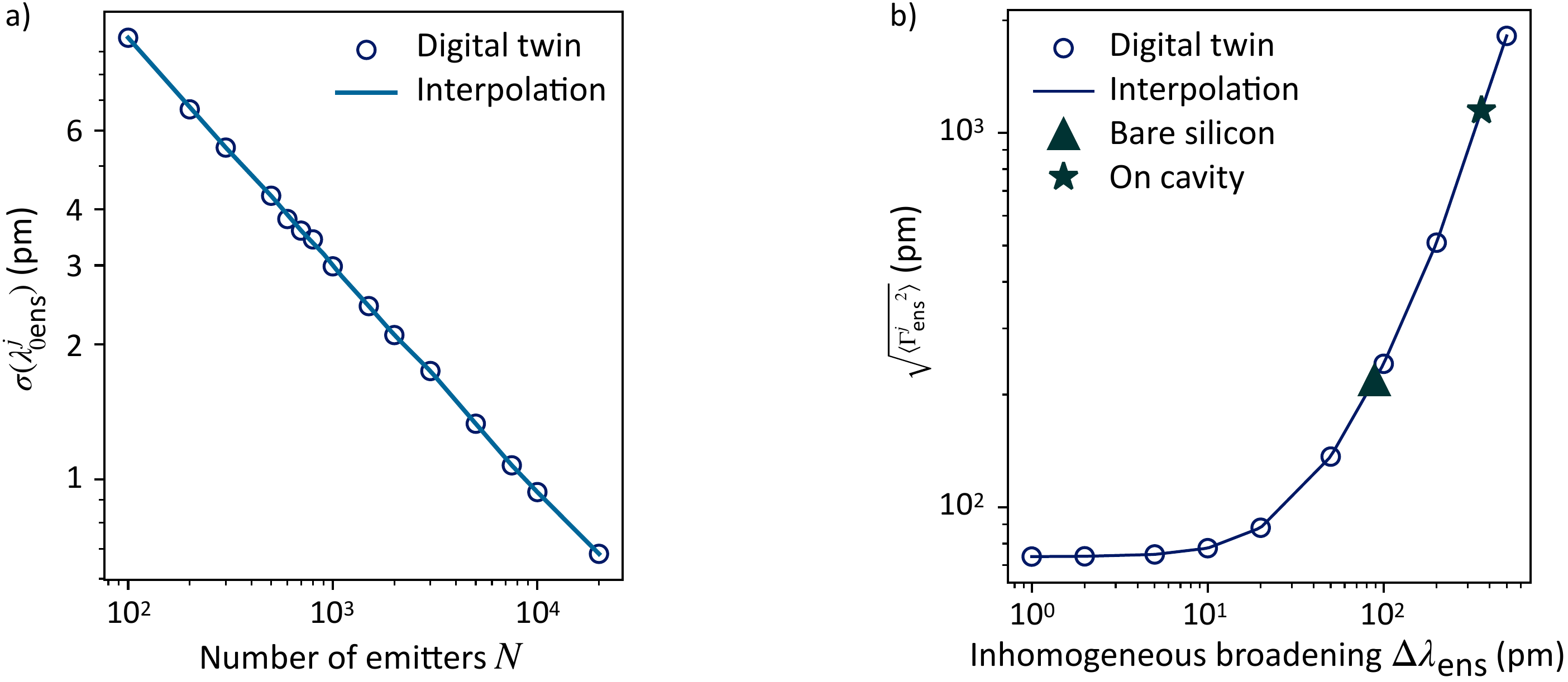}
    \caption{\textbf{Statistical spectral analysis on ensembles of G-centers.} a) Standard deviation of ensemble central wavelengths versus the number of created emitters. As the number of emitters increases, the standard deviation decreases following Eq.~\ref{sigma}. The solid line represents the interpolation of the digital twin data (circles). b) FWHM of an ensemble lineshape versus the inhomogeneous broadening. By extracting $\sqrt{\langle {\Gamma^j_{\mathrm{ens}}}^2 \rangle}$ from the experimental data, we can deduce the corresponding inhomogeneous broadening for both the bare silicon and on-cavity annealing cases from the interpolation (solid line) of the digital twin data (circles).}
    \label{sfig:inhomogenousbroadening}
\end{figure}

Therefore, for a single ensemble spectrum measurement, we draw a sample from $\lambda_{0\mathrm{ens}}^j\sim\mathcal{N}(\lambda_0, {\Delta \lambda_{\mathrm{ens}}}^2)$ and a sample from $\Gamma_{\mathrm{ens}}^j\sim\mathcal{N}(\Gamma_0, {\Delta \Gamma_{\mathrm{ens}}}^2)$.
For repeated measurements of the spectra, we expect 
\begin{align}
\langle \lambda_{0\mathrm{ens}}^j\rangle &\rightarrow \lambda_0, \label{lambda} \\
{\sigma (\lambda_{0\mathrm{ens}}^j)} &\rightarrow \Delta \lambda/\sqrt{N}, \label{sigma} \\
\langle {\Gamma^j_{\mathrm{ens}}}^2\rangle &\rightarrow \Delta \lambda^2+\sigma_0^2, \label{Gamma}
\end{align}
where $\langle\cdot\rangle$ indicates the ensemble average and $\sigma(\cdot)$ is the standard deviation. Eq.~\ref{sigma} shows that the spread of the fitted central wavelength decreases with the total number of emitters, while Eq.~\ref{Gamma} suggests that the FWHM of the ensemble spectra tends to the inhomogeneous distribution of the underlying emitter distribution convolved with the IRF. 

In this model, we further assume that the measured PL signal at each wavelength $I(\lambda)$ follows a Poissonian distribution centered around $\mathrm{PL_{ens}}(\lambda)$ (i.e. we are adding photon shot noise to our model): 
\begin{equation}
    I(\lambda)\sim \mathrm{Poisson}\left( \mathrm{PL_{ens}}(\lambda) \right).
\end{equation}
The digital twin takes in given arbitrary values of $\Delta \lambda$, spectrometer camera pixelization and IRF, and produces a spectrum with random sampling due to emitter non-uniformity and photon statistics. The resulting instances of distributions are then used to compute the values $\sqrt{\langle {\Gamma^j_{\mathrm{ens}}}^2 \rangle}$ and $\sigma (\lambda_{0\mathrm{ens}}^j)$ as shown in Fig.~\ref{sfig:inhomogenousbroadening}. From experiments, we extract the values $\sigma (\lambda_{0\mathrm{ens}}^j)$ and $\sqrt{\langle {\Gamma^j_{\mathrm{ens}}}^2 \rangle}$, which are then considered in the digital twin model. By tuning the unknown parameters $N$ and $\Delta \lambda$, the digital twin can reproduce the experimental results without assumed or extracted parameters. We determine the number of emitters to be lower bounded by $\sim~(246\pm90)$, and the inhomogeneous broadening to be ${\Delta \lambda_{\mathrm{ens}}}=(88.5\pm1.5)~\mathrm{pm}$ for bare silicon. In comparison, the value $\sqrt{\langle {\Gamma^j_{\mathrm{ens}}}^2 \rangle}$ for ensembles on cavities 
is $1.1~\mathrm{nm}$, resulted from an inhomogeneous broadening of ${\Delta \lambda_{\mathrm{ens}}}=360~\mathrm{pm}$. This number is larger than that of bare silicon, likely due to local strain created by nanofabrication.

\subsection{Digital twin for emitters on and off optical cavities}
\label{DTcavity}
In this section, we extend the transient PL digital twin modeling to the case where annealing is performed on nanopatterned structures such as photonic crystal optical cavities, rather than just bulk silicon. We therefore consider two additional Hamiltonian terms: the cavity Hamiltonian 

\begin{figure}[t]
    \centering
    \includegraphics[width=0.42\linewidth]{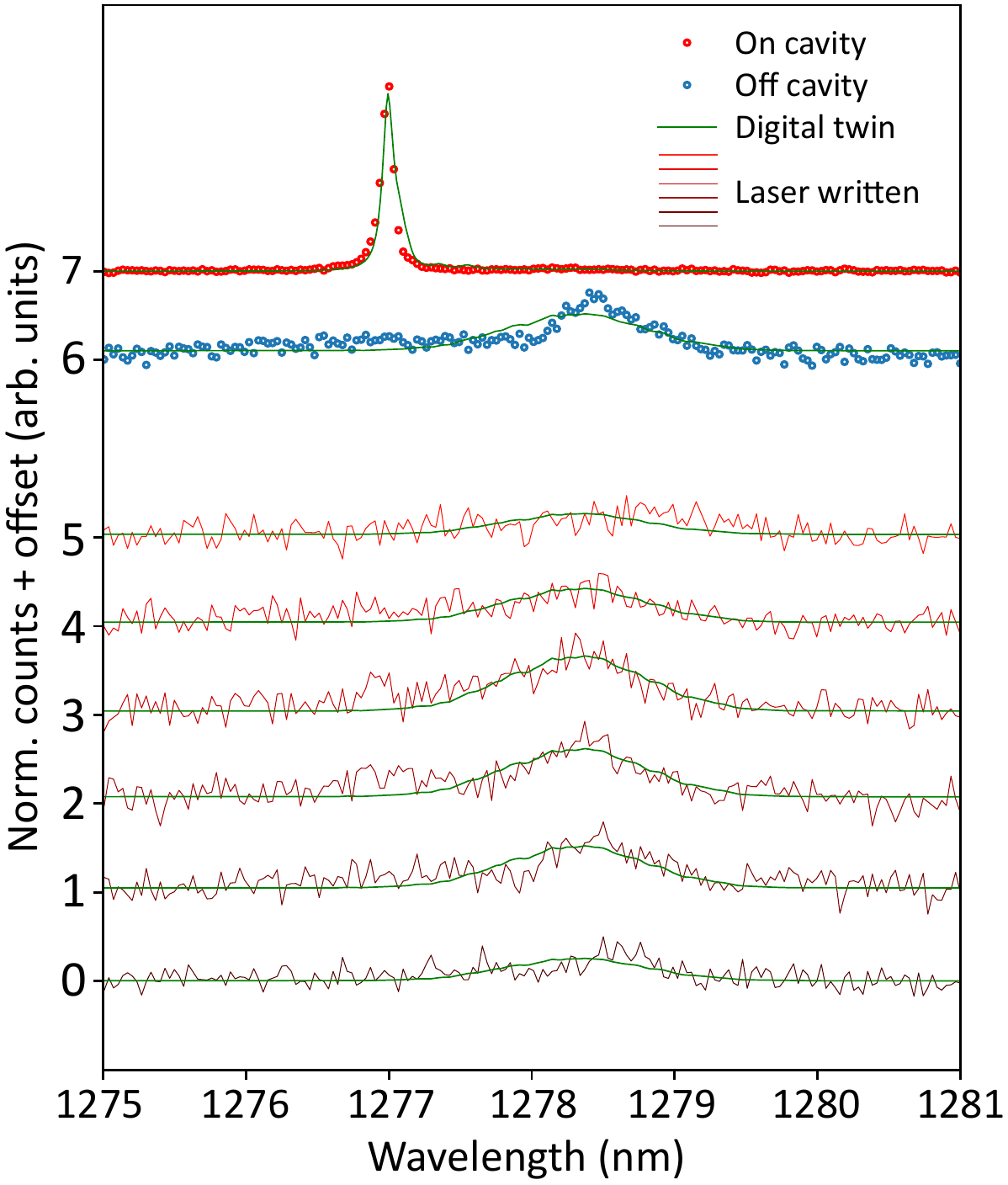}
    \caption{\textbf{Digital twin modeling for Figs.~\ref{fig4}c, d.} From the bottom: the first six curves are the spectra of laser-written emitters along a line off the cavity already shown in Fig.~\ref{fig4}d (note that only spectra with a visible emission feature are reported here); the blue circles represent the spectrum of emitters (not annealed) excited on the blob to the right of the cavity visible in Fig.~\ref{fig4}c. The red circles represent the spectrum of an emitter (not annealed) excited on the cavity. In all these cases --- whether annealing is performed, and whether a cavity is present --- the digital twin (solid green line) accurately reproduces the experimental results.}
    \label{sFig:fig4DTPanelsB}
\end{figure}

\begin{equation}
H_{\mathrm{cavity}}=\omega_{\mathrm{c}} 
a^\dagger a,
\end{equation}
where $a^\dagger$ is the photon creation operator, $a$ the photon annihilation operator, and $\omega_{\mathrm{c}}$ the cavity resonance frequency, and the atom-cavity coupling Hamiltonian 
\begin{equation}
    H_{\mathrm{atom-cavity}} = \Omega (a\sigma^++a^\dagger\sigma^-),
\end{equation}
where $\Omega$ is the atom-cavity coupling constant, $\sigma^+=|\mathrm{ES}\rangle\langle\mathrm{GS}|$ and $\sigma^{-}=|\mathrm{GS}\rangle\langle\mathrm{ES}|$.
Additionally, the cavity is lossy and contributes a collapse operator $\sqrt{\kappa} a$, where $\kappa$ is the cavity decay rate and is extracted from the experimentally measured quality factor $Q$ of the cavity. Specifically, $Q=\lambda_{\mathrm{c}}/\Delta \lambda_{\mathrm{c}}=c/(\lambda_{\mathrm{c}}\kappa)$, where $\lambda_{\mathrm{c}}$ and $\Delta \lambda_{\mathrm{c}}$ are the central wavelength and FWHM of the cavity resonance profile, respectively. The electric field created by the cavity mode and emitters' emission is of the form~\cite{englund2010deterministic} 

\begin{equation}
    E = \sqrt{\kappa}a\hat{e}_{c} + \sum_{i=0}^{12} \sqrt{\xi_i\Gamma_a}\sigma^-\hat{e}_{G},
\end{equation}
where $\hat{e}_{c}$ and $\hat{e}_{G}$ describe the spatial profile of the cavity mode and G-center emission. The PL spectrum is derived from the Fourier transform of the one-time correlation function, computed via the QuTiP package:
\begin{equation}
    L(\omega) \propto \int dt E^\dagger E(0)\langle \rho(t)\rho(0)\rangle e^{i\omega t},
\end{equation}
 where $\omega$ is the angular frequency of the emitted light. $L(\omega)$ consists of multiple terms, including the emitters' emission 
\begin{equation}
    L_{G}(\omega)\propto\sum_{i=0}^{12} {\xi_i\Gamma_a}\int dt\langle\sigma^+\sigma^-\rangle e^{i\omega t},
\end{equation}
the cavity emission,
\begin{equation}
    L_{c}(\omega)\propto{\kappa}\int dt\langle a^\dagger a\rangle e^{i\omega t},
\end{equation}
as well as the cross terms

\begin{figure}[t]
    \centering
    \includegraphics[width=1\linewidth]{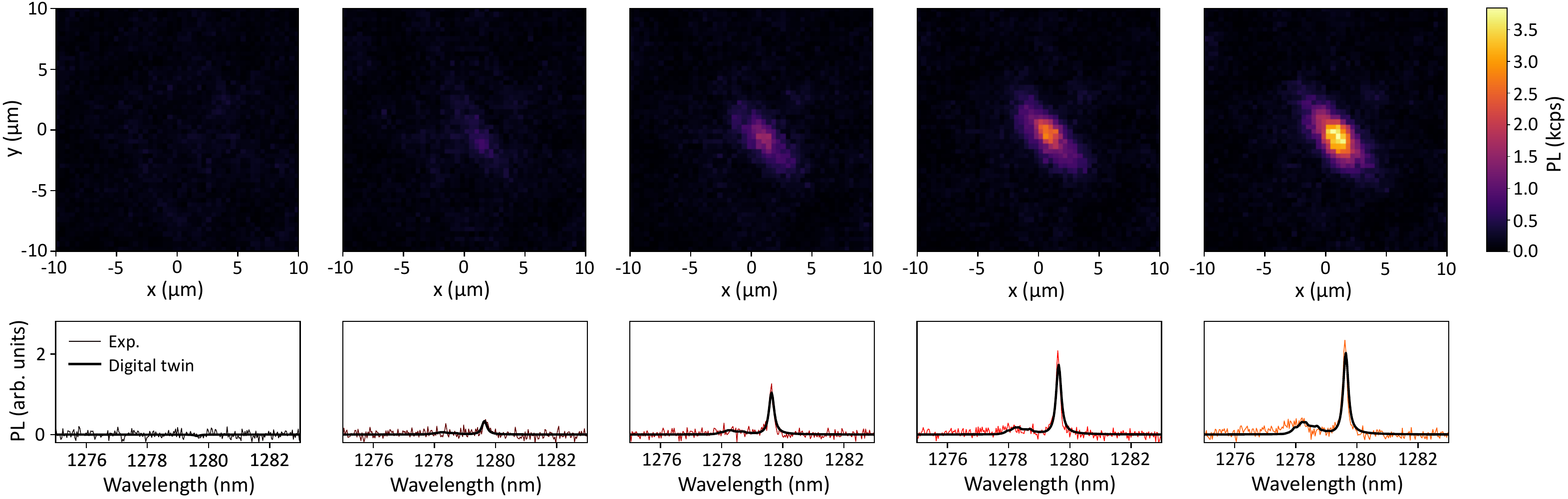}
    \caption{\textbf{Digital twin modeling for Figs.~\ref{fig4}e, f.} The top row shows the PL maps of the cavity reported in Fig.~\ref{fig4}e under increasing annealing repetitions. The bottom row displays the corresponding spectra (reported all together in Fig.~\ref{fig4}f). The progressive appearance of a small peak is observed next to a higher peak, which corresponds to emission into the cavity mode. We attribute the small peak to the ZPL of annealed emitters, and the bigger peak to the cavity-enhancement of their phonon sidebands. The digital twin (solid black line) is able to accurately reproduce the spectral properties of both the annealed emitters and their cavity enhancement.}
    \label{sFig:fig4DTPanelsD}
\end{figure}

\begin{equation}
    L_{c-G}(\omega)\propto\sqrt{\xi_i \kappa}\int dt \mathcal{R}\left(\langle \sigma^+a + \sigma^-a^\dagger\rangle e^{i\omega t}\right),
\end{equation}
where $\mathcal{R}$ denotes the real part. The final spectrum is the sum of these terms with respective weighting determined by the collection efficiency of each spatial modes, which are uncertain in our experiments. We thus use three fitting parameters $C_1$, $C_2$, and $C_3$ to account for this uncertainty, leading to the function $L(\omega)=C_1 L_{G}(\omega) + C_2L_{c-G}(\omega)+ C_3L_{c}(\omega)$.

For all emitters presented in Figs.~\ref{fig4}c-f in the main text, we find a ZPL of $\lambda_a=1278.31~\mathrm{nm}$ and a Gaussian distribution of emitters with a standard deviation of $486~\mathrm{pm}$. The cavity displayed in panel~\ref{fig4}c has a resonance of $\lambda_{\mathrm{c}}=1276.99~\mathrm{nm}$ and a quality factor $Q=14148$, while the one reported in panel~\ref{fig4}e features $\lambda_{\mathrm{c}}=1279.63 ~\mathrm{nm}$ and $Q=7075$. The digital twin correctly accounts for the spectral features of the three classes of emitters shown in Fig.~\ref{sFig:fig4DTPanelsB}. With reference to Fig.~\ref{fig4}c and d, the first class comprises laser-written emitters along a line off the cavity center on a photonic crystal structure, showing a broad ensemble resonance at $\lambda_a=1278.31~\mathrm{nm}$ (the six lines counting from the bottom in Fig.~\ref{sFig:fig4DTPanelsB}). The second class includes emitters present before laser writing also showing ensemble resonance at $\lambda_a=1278.31~\mathrm{nm}$ (blue dots in Fig.~\ref{sFig:fig4DTPanelsB}, corresponding to a measurement done on the blob to the right of cavity visible in Fig.~\ref{fig4}c). The third class is composed of emitters at the center of the cavity and emitting into the cavity mode at $1.32~\mathrm{nm}$ away from the atomic emission (red dots in Fig.~\ref{sFig:fig4DTPanelsB}). The solid green curve represents the digital twin predictions for each of these cases, and shows agreement with all the different experimental conditions mentioned above.

Our digital twin is a unified model able to reproduce the results of emitters annealed on the cavity center as well, as shown in Fig.~\ref{sFig:fig4DTPanelsD}. The bottom row reports the spectra already shown in Fig.~\ref{fig4}f, fitted using the digital twin model. As more annealing repetitions are performed on the cavity center, a small peak starts to appear with a central wavelength of $1278.31~\mathrm{nm}$ next to a higher peak aligned with the cavity resonance. The smaller peak indicates the creation of an ensemble of G-centers, and the presence of the cavity likely enhances their phonon sidebands at its resonance wavelength. As the number of annealing rounds increases, the PL emitted from the cavity also intensifies, as visible in the top row of Fig.~\ref{sFig:fig4DTPanelsD}. 

\appendix*

\end{document}